\documentclass[twocolumn]{aastex631}

\usepackage{color}
\usepackage{url}
\shorttitle{X-ray polarization in BL Lac}
\shortauthors{Middei et al.}

\begin{document}

\title{X-ray Polarization Observations of BL Lacertae}



%
%
%
%
%
%


\author[0000-0001-9815-9092]{Riccardo Middei}
\correspondingauthor{Riccardo Middei}
\email{riccardo.middei@ssdc.asi.it}
\affiliation{Space Science Data Center, Agenzia Spaziale Italiana, Via del Politecnico snc, 00133 Roma, Italy}
\affiliation{INAF Osservatorio Astronomico di Roma, Via Frascati 33, 00078 Monte Porzio Catone (RM), Italy}

\author[0000-0000-0000-0000]{Ioannis Liodakis}
\affiliation{Finnish Centre for Astronomy with ESO, FI-20014 University of Turku, Finland}

\author[0000-0000-0000-0000]{Matteo Perri}
\affiliation{Space Science Data Center, Agenzia Spaziale Italiana, Via del Politecnico snc, 00133 Roma, Italy}
\affiliation{INAF Osservatorio Astronomico di Roma, Via Frascati 33, 00078 Monte Porzio Catone (RM), Italy}

\author[0000-0000-0000-0000]{Simonetta Puccetti}
\affiliation{Space Science Data Center, Agenzia Spaziale Italiana, Via del Politecnico snc, 00133 Roma, Italy}


\author[0000-0001-7150-9638]{Elisabetta Cavazzuti}
\affiliation{ASI - Agenzia Spaziale Italiana, Via del Politecnico snc, 00133 Roma, Italy}

\author[0000-0000-0000-0000]{Laura Di Gesu}
\affiliation{ASI - Agenzia Spaziale Italiana, Via del Politecnico snc, 00133 Roma, Italy}

\author[0000-0003-4420-2838]{Steven R. Ehlert}
\affiliation{NASA Marshall Space Flight Center, Huntsville, AL 35812, USA}

\author{Grzegorz Madejski}
\affiliation{Department of Physics and Kavli Institute for Particle Astrophysics and Cosmology, Stanford University, Stanford, California 94305, USA}

\author[0000-0001-7396-3332]{Alan P. Marscher}
\affiliation{Institute for Astrophysical Research, Boston University, 725 Commonwealth Avenue, Boston, MA 02215, USA}

\author[0000-0002-6492-1293]{Herman L. Marshall}
\affiliation{MIT Kavli Institute for Astrophysics and Space Research, Massachusetts Institute of Technology, 77 Massachusetts Avenue, Cambridge, MA 02139, USA}

% ...

\begin{abstract}
Blazars are a class of jet-dominated active galactic nuclei with a typical double-humped spectral energy distribution. It is of common consensus the Synchrotron emission to be responsible for the low frequency peak, while the origin of the high frequency hump is still debated. The analysis of X-rays and their polarization can provide a valuable tool to understand the physical mechanisms responsible for the origin of high-energy emission of blazars. We report the first observations of BL Lacertae performed with the Imaging X-ray Polarimetry Explorer ({IXPE}), from which an upper limit to the polarization degree $\Pi_X<$12.6\% was found in the 2-8~keV band. We contemporaneously measured the polarization in radio, infrared, and optical wavelengths. Our multiwavelength polarization analysis disfavors a significant contribution of proton synchrotron radiation to the X-ray emission at these epochs. Instead, it supports a leptonic origin for the X-ray emission in BL Lac.
\end{abstract}


\section{Introduction} \label{sec:intro}
Observations of astrophysical jets from supermassive black holes offer unique opportunities to study energetic multi-waveband emission processes in the Universe \citep[see e.g.,][]{Blandford2019}. Blazars are a subclass of active galactic nuclei (AGN) whose jets are aligned within a few degrees of the line of sight. They are often characterized by superluminal motion of bright knots in their jets, and their emission, which is relativistically Doppler-boosted, exhibits extreme variability across the electromagnetic spectrum \citep[e.g.,][]{Hovatta2019}. Their radio and optical emission is significantly linearly polarized \citep[e.g.,][]{Agudo2018,Blinov2021}, which is attributed to synchrotron radiation from relativistic electrons in the jet. The origin of the broad keV-to-TeV emission component is a matter of current debate. Most studies interpret the high-energy photons as the result of Compton scattering. The seed photons could originate from either the jet's synchrotron radiation (synchrotron self-Compton; SSC) or from external radiation fields (external Compton; EC). 
This scenario has been supported by spectral energy distribution (SED, e.g., \citealp{Abdo2011-II}) modeling, energetic considerations \citep{Zdziarski2015,Liodakis2020-II}, observations of flux variations that are correlated across the wavebands (e.g., \citealp{Agudo2011,Agudo2011bis,Liodakis2018,Liodakis2019-II}), and low or even undetected in radio/optical circular polarization \citep{Wardle1998,Liodakis2022}. 
However, scenarios invoking proton-initiated emission \citep[synchrotron radiation by relativistic protons, and/or emission processes associated with cascades of leptons produced by photo-hadronic processes, e.g.,][]{IceCube2018} have not been definitively excluded. Typically, leptonic models have been more successful in modeling low synchrotron peaked blazars (LBL, $\nu_{syn}<10^{14}~{\rm Hz}$), while hadronic models are often favored for high-synchrotron-peak sources ($\nu_{syn}>10^{15}~{\rm Hz}$, e.g., \citealp{boettcher2013,Cerruti2015,Cerruti2017}).

Measurements of X-ray polarization can be used to test high-energy emission processes and particle acceleration in jets \citep[e.g.,][]{Zhang2013,Liodakis2019,Tavecchio2018}. Starting in January 2022, the Imaging X-ray Polarimetry Explorer \citep[{ IXPE},][]{weisskopf2010,Weisskopf2016,Weisskopf2022} has been carrying out such measurements. Detection by {IXPE} of high-synchrotron-peak sources like Mrk 501 and Mrk 421 \citep{Liodakis2022-Mrk501,DiGesu2022-Mrk421} has revealed stronger polarization at X-rays than at longer wavelengths. This is consistent with emission by high-energy electrons that are accelerated at a shock front with partially-ordered magnetic fields, after which they are advected to regions with more turbulent fields. 

Here we report the first X-ray polarimetric observations of a LBL blazar, the
eponymous BL Lacertae \citep[BL Lac, $z=0.0686$,][]{Vermeulen1995}. The X-ray emission of BL Lac is highly variable, with an average flux of $\rm F_{\rm 2-10~keV}\sim 1 \times 10^{-11}$ erg cm$^{-2}$ s$^{-1}$  \citep[e.g.][]{Wehrle2016,Giommi2021,Sahakyan2022,Middei2022blaz}. 
Moreover, BL Lac is a very high energy (VHE) emitting source as it is the 14th brightest AGN at $\gamma$-ray energies listed in the Fermi 4LAC catalog \citep{Ajello2020} and among the few LBL sources detected in TeV $\gamma$-rays showing fast, even down to $\sim$hourly timescales, flux variability  \citep{Albert2007,Arlen2013}.
Moreover, BL Lac has been the focus of a large number of multi-wavelength and polarization studies (e.g., \citealp{Raiteri2013,Blinov2015,Blinov2018,Weaver2020,Casadio2021}).

This paper is organized as follows. We describe the {IXPE} observations and X-ray data processing and analysis in section \ref{sec:xray+pol} and our contemporaneous observing campaign at radio, infrared, and optical wavelengths in section \ref{sec:Multi_data}. Finally, we discuss and interpret our results in section \ref{sec:disc_conc}. A standard  $\Lambda$CDM cosmology with H$_{\rm 0}$ = 70 km s$^{-1}$ Mpc$^{-1}$, $\Omega_{\rm m}$ = 0.27 and $\Omega_{\lambda}$ = 0.73 is adopted throughout this work. Errors quoted in text and in
plots correspond to 1$\sigma$ uncertainties ($\Delta \chi^{2}=1$ for 1 parameter of interest). All upper limits related to IXPE observations are quoted at 99\% confidence,
corresponding to $\Delta \chi^{2} =6.635$ for
1 parameter of interest.

\section{X-ray Spectra and polarization observations}\label{sec:xray+pol}

BL Lac was observed with the three Detector Units (DUs) of {IXPE} during 2022 May 6--14  for a net exposure of 390 ks. The second observation was performed 2022 July 9--11 for a net exposure of $\sim$116 ks. Quasi-simultaneously with the first {IXPE} observation, BL Lac was observed by the Nuclear Spectroscopic Telescope Array \citep[NuSTAR,][]{Harrison2013}, with a $\sim$25 ks exposure, and with the {EPIC-pn} \citep{Struder2001A&A...365L..18S} camera on board {XMM-Newton} \citep{Jansen2001}. In addition, another {XMM-Newton} observation was taken simultaneously with the second {IXPE} observation of BL Lac. In Appendix:~\ref{app:X-ray_obs}, we report the details on the data reduction of the {IXPE, XMM-Newton} and {NuSTAR} data. The X-Ray Telescope \cite[XRT,][]{Burrows2005} on the {Neil Gehrels Swift Observatory} ({Swift}) monitored the blazar from 2022 May until July. Details on the data reduction and the results of these observations are provided in Appendix:~\ref{sec:swift_obs}.

\subsection{X-ray spectral analysis}\label{subsec:xray_analysis}
\indent We combine the {IXPE} Stokes $I$ (i.e., total flux density), {XMM-Newton}, and (only for the first 
exposure) {NuSTAR} data to determine the X-ray spectrum in the 0.5-79 keV energy range. We first attempt to fit the data with a simple model of a single power-law continuum with photoelectric absorption exceeding that from gas in our Galaxy. In fact, various studies \citep[e.g.,][]{Bania1991,Madejski1999} have reported on the presence of neutral absorption in this source, invoking the presence of molecular clouds along the line of sight to BL Lac. We thus have fit the column density requiring a consistent value for both observations 1 and 2.
and {NuSTAR}) and observation 2 ({IXPE} and {XMM-Newton}). These steps led to a good fit with $\chi^2$ statistic $\chi^2$/d.o.f=1234/1112. The column density derived exceeds the Galactic value as expected and the {XMM-Newton} observations show bump-like residuals around 0.7 keV, which we infer as being due to an additional spectral component. We speculate that it represents emission from hot diffuse plasma, which we include as an {\it apec} model in {\it XSPEC}. We fit the temperature and the normalization of this {\it apec} component while requiring its temperature and normalization to be consistent between the two datasets. This step led us to a satisfactory spectral fit ($\chi^2$/d.o.f.=1169/1110). We attempted to replace the apec component with a single Gaussian centered at 0.7 keV; however, this returned a worse fit with $\Delta\chi^2$=+31 for the same number of degrees of freedom. Finally, the present dataset does not support the presence of a Synchrotron component at soft X-ray energies, which would be the tail of the low energy hump of the SED. Replacing the {\it apec} model with such a steep power-law component degrades the quality of the fit to the data ($\Delta\chi^2$=+52 with two additional free parameters).

The cross-normalization constants between {IXPE} and {XMM-Newton} were consistent with unity within $\sim$10\%.  The difference between the {IXPE} and {NuSTAR} flux normalizations was $\sim$30\%, although this could be mainly ascribed to the flux level of BL Lac being higher during the {NuSTAR} pointing than the average during the {IXPE} exposure.

Based on our model fits, the X-ray spectrum of BL Lac beyond $\sim 2$ keV was characterized by a power law with photon index $\Gamma=1.74\pm$0.01 and 1.87$\pm$0.06 for observations 1 and 2, respectively. The source was in a higher flux level during the later epoch. The absorbing column density is found to be N$_{\rm H}$=2.60$\pm$0.05$\times10^{21}$ cm$^{-2}$ and this value is in perfect agreement with the extensive analysis performed by \citet{Weaver2020}. At low energies, below the IXPE bandpass, the spectra are consistent with emission from hot gas ($kT=0.38\pm$0.04 keV). However, the physical origin of such an additional soft component is unknown and requires additional observations to determine.

\subsection{Spectro-polarimetric X-ray analysis}\label{subsec:spec_pol}
\begin{figure*}
    \centering
     \includegraphics[width=0.45\textwidth]{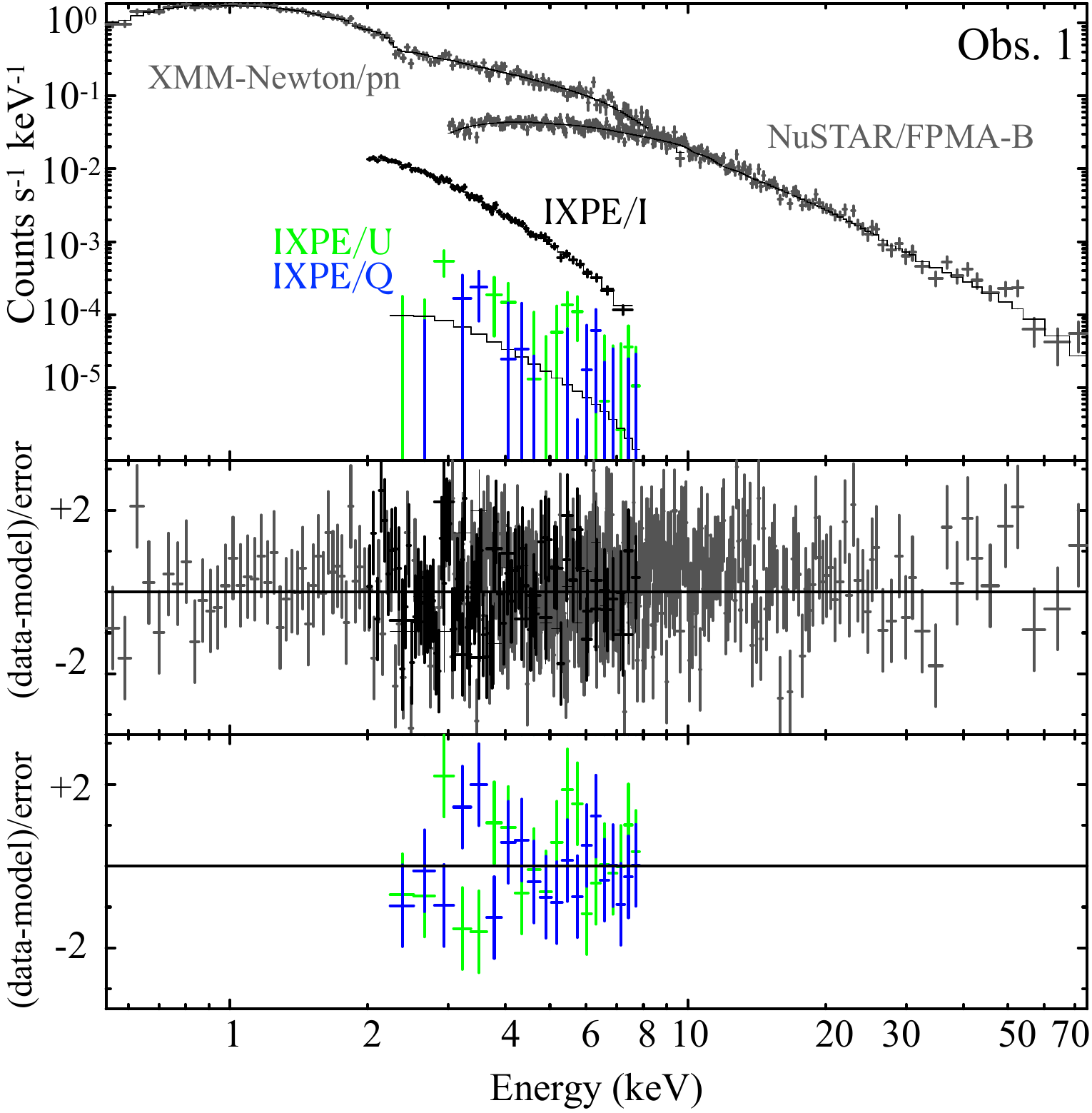}
    \includegraphics[width=0.45\textwidth]{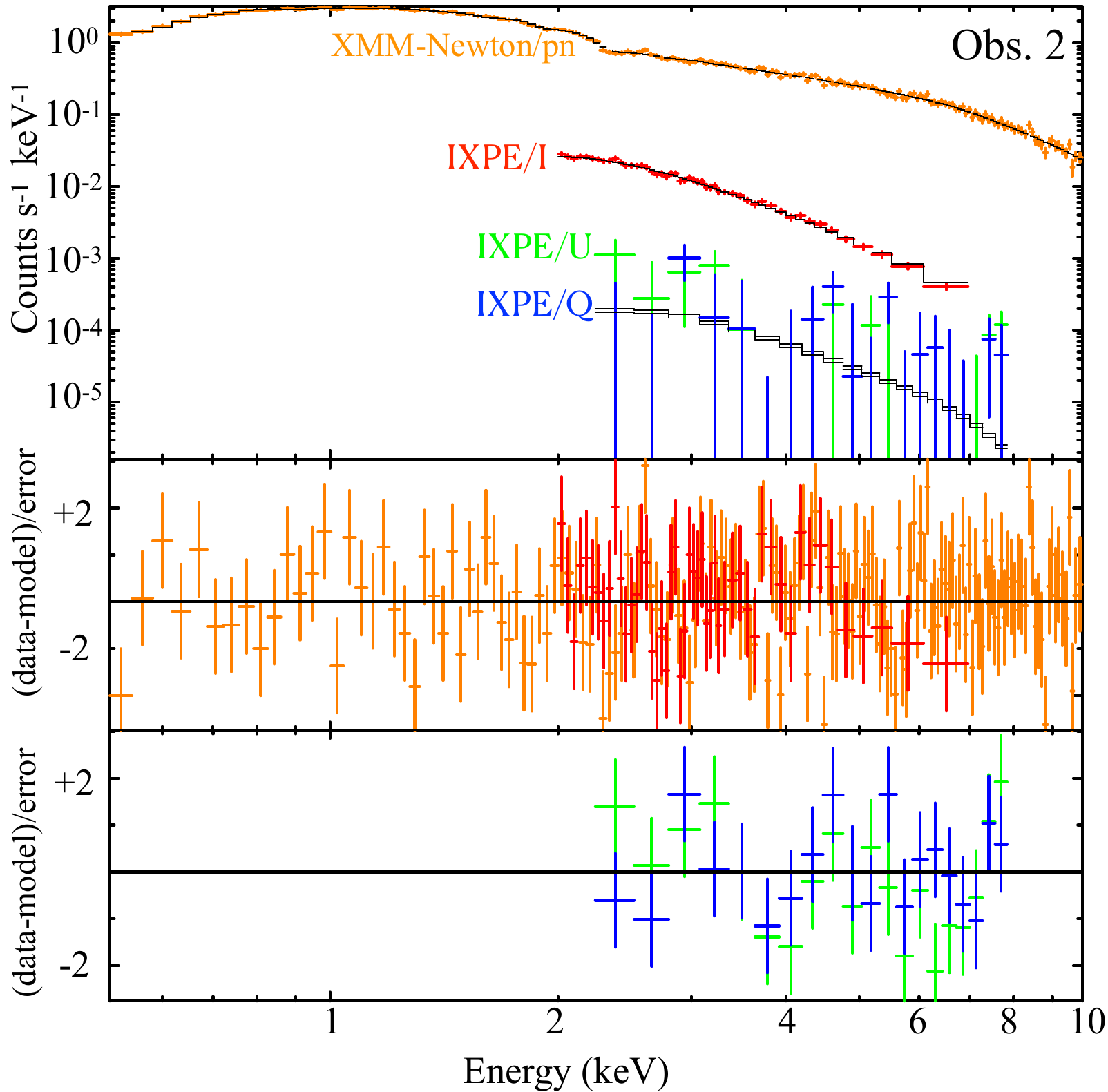}
    \includegraphics[width=0.46\textwidth]{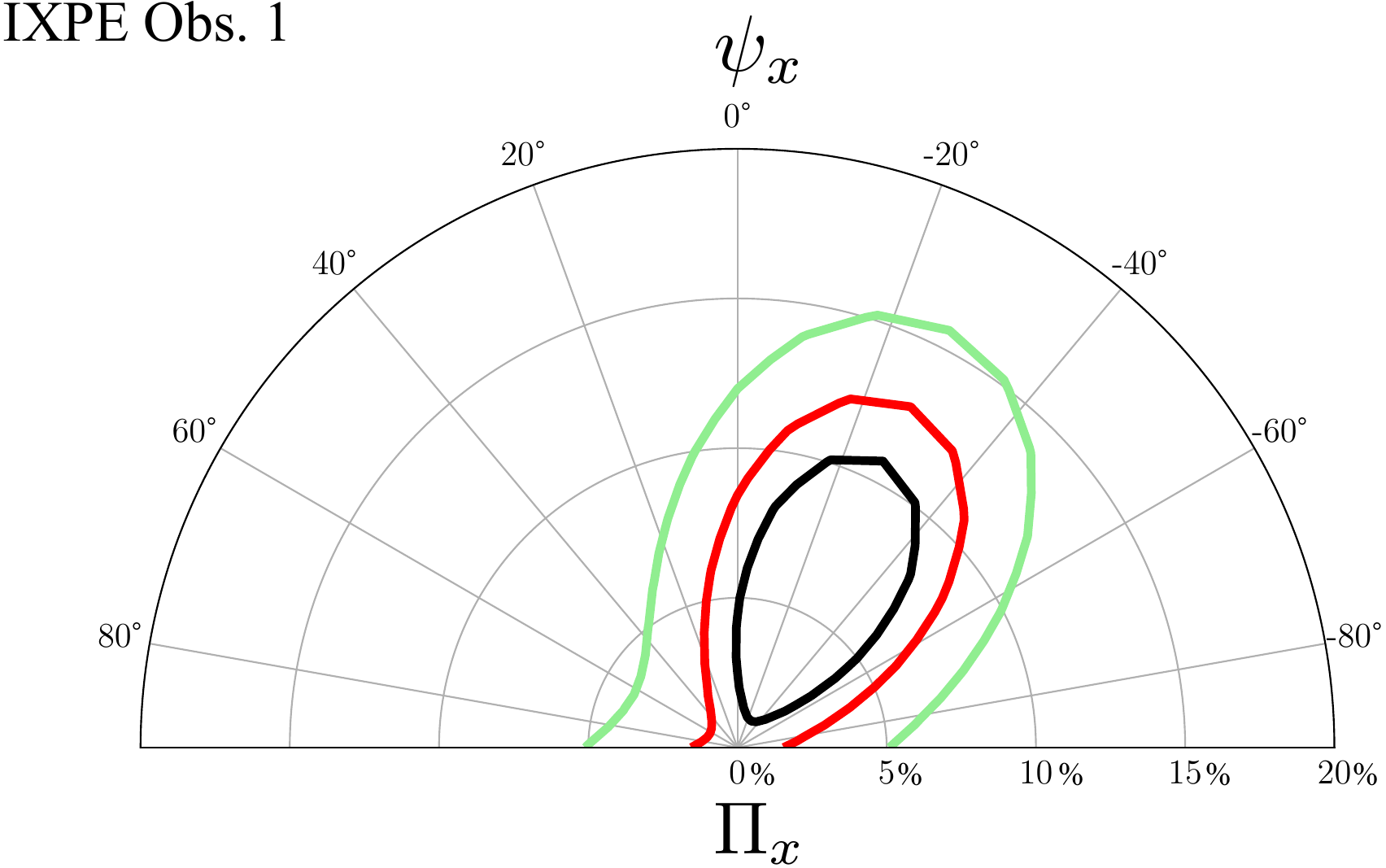}
    \includegraphics[width=0.46\textwidth]{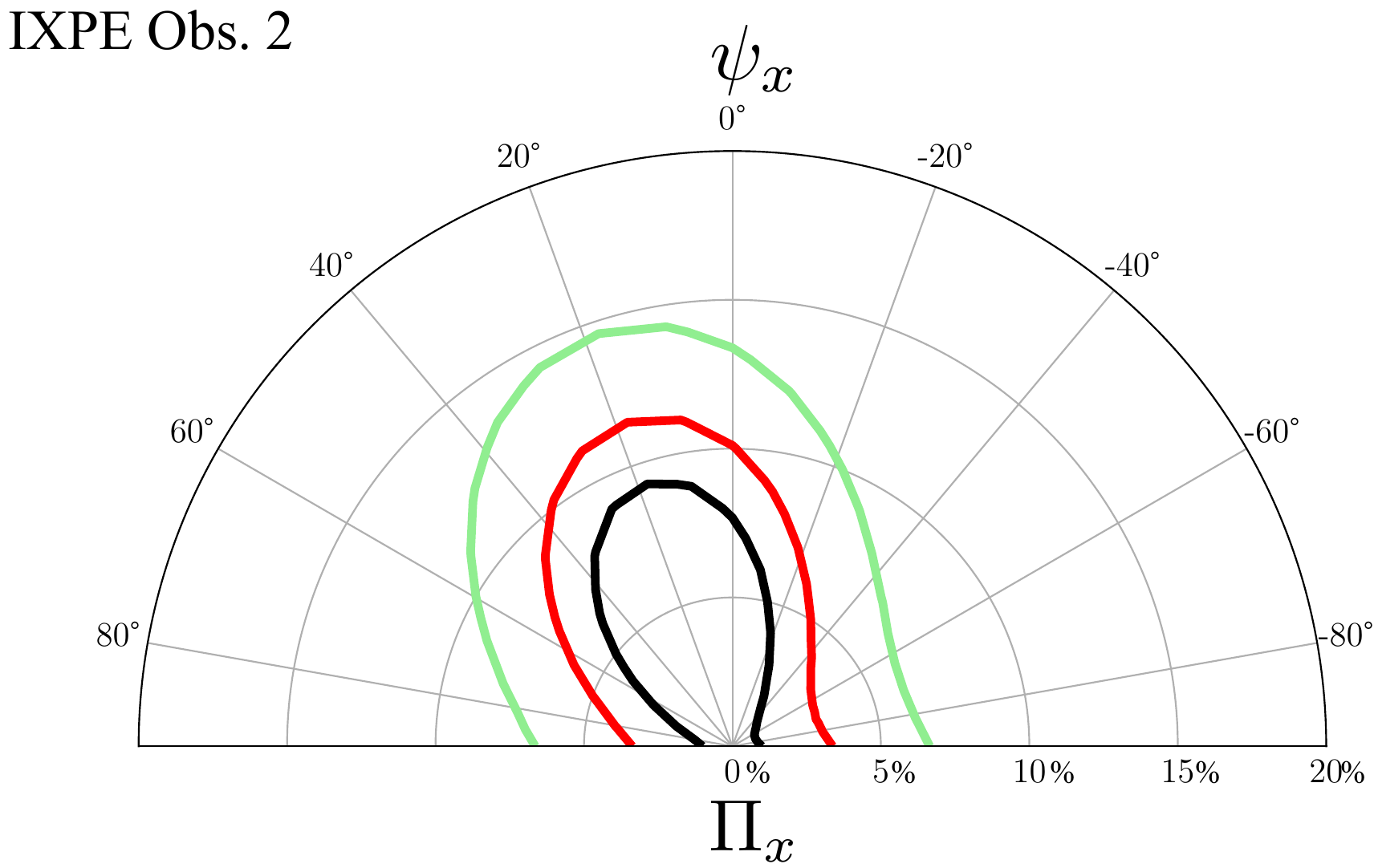}
    \caption{Top panels: Best-fit to the May X-ray dataset (on the left) and the IXPE-XMM-Newton data taken during July 2022 (right plots). In the middle and bottom sub-panels, we report the residuals to the I  and the $U$ and $Q$ spectra, respectively.
    {\it Bottom plots:} Confidence regions of the model fits for the polarization angle and degree for the two observations.}
    \label{xspecpol}
\end{figure*}

We searched for X-ray polarization from BL Lac by performing a spectro-polarimetric fit of the $I$, $Q$ and $U$ Stokes spectra over the two {IXPE} exposures. To better constrain the spectral shape of BL Lac we performed the spectro-polarimetric analysis also including the XMM-Newton and NuSTAR data. Similarly to the fit to the $I$ Stokes spectra, we fitted simultaneously the Galactic column density and the {\it apec} component, while the photon index and the normalization of the power-law were computed for each observation.
We then accounted for the polarimetric information encoded in the $Q$ and $U$ Stokes parameters multiplying the power-law with {\it polconst}, i.e. a {\it XSPEC} model that assumes the polarization signal to be constant across the {IXPE} energy range. Thus, the final model consists of $tbabs \times const \times (apec + polconst \times powerlaw$), in {\it XSPEC} notation. The constant accounts for the inter-calibration among the different detector units.
We first fit separately the two IXPE observations with the polarization degree and angle free to vary between exposures. This procedure leads to a best-fit of $\chi^2$=1469 for 1388 d.o.f., and provides two upper limits for the polarization degree: $\rm \Pi_X=<14.2\%$ and $\rm \Pi_X<12.6\%$ (at 99\% confidence level) for Obs.\ 1 and Obs.\ 2, respectively. In Figure~\ref{xspecpol} we display the fit to the two {IXPE} exposures and the corresponding confidence regions. In Table~\ref{ixpefit} we report the best-fit values corresponding to the analysis of the {IXPE} observations.
Then we tested a scenario where both $\Pi_X$ and $\psi_X$ remained unchanged between observations. Although blazars typically vary on shorter timescales than two months, this test is motivated by the fact that the polarization angle in the mm/radio energy range $\psi_R$ (see Section 3), i.e., the seed photons in case of synchrotron self-Compton emission, is consistent within uncertainties between the two observations (see Appendix \ref{sec:multi_obs}). Moreover, as found for hadronic models including polarization, X-ray polarization is expected to be less variable than at optical wavelengths \citep{Zhang2016}. 
This simple test yields a compatible fit ($\chi^2$/d.o.f.=1473/1390), with the spectral parameters being consistent with those quoted in Table~\ref{ixpefit}. Also in this case, we obtain only an upper limit to the polarization degree, $\Pi_X<9.6\%$, and the polarization angle is unconstrained. We subsequently set the polarization degree to be the same between the observations, but allow the polarization angle to vary. This attempt, which yields only a compatible fit statistic, gives $\rm \Pi_X<11.1\%$ with no information on the polarization angle. We also tested the opposite scenario in which the polarization angle is constant between the two {IXPE} exposures and $\Pi_X$ varies. Such a test, to which corresponds an equivalent fit statistic, led us to an unconstrained X-ray polarization angle and upper limits for the polarization degrees of $\Pi_X<14.2\%$ and $\Pi_X<11.7\%$ for the two observations.

\begin{table}
\setlength{\tabcolsep}{.1pt}
\centering
	\caption{\small{Best-fit parameters for the two {IXPE} observations.  The power-law normalization is in units of $\times$10$^{-3}$ photons keV$^{-1}$ cm$^{-2}$ s$^{-1}$, the apec component has a normalization of $\times$10$^{-4}$, and fluxes are $\times$10$^{-11}$ erg cm$^{-2}$ s$^{-1}$}\label{ixpefit}. Errors accounting for the polarimetric information refer to 99\% confidence interval for 1 parameter of interest.}
	\begin{tabular}{c c c c}
	\hline
	Model & Component & Obs. 1 & Obs. 2 \\
	\hline
    polconst& $\rm \Pi_X$ &$<14.2\%$ & $<12.6\%$\\
     & $\rm \psi_X$ &-&-\\
    tbabs& N$_{\rm H}\dagger$ &2.60$\pm$0.05 &\\
    apec& kT (keV)$\dagger$ &0.38$\pm$0.04 &\\
    & normalization &3.1$\pm$0.1 &\\
    powerlaw& $\Gamma$ & 1.74$\pm0.01$& 1.87$\pm$0.06\\
    & Norm &$2.74\pm0.05$ &$5.5\pm0.1$ \\
    \hline
    $\rm F_{\rm 2-8~keV}$&& $0.96\pm0.03$(0.05)& $1.56\pm0.06$(0.09)\\
    \hline
		\end{tabular}
	\end{table}

\section{Multi-wavelength observations} \label{sec:Multi_data}

During the {IXPE} observations, a number of telescopes provided multi-wavelength polarization coverage: the Atacama Large Millimeter Array (ALMA), AZT-8 (Crimean Astrophysical Observatory, 70 cm diameter), Calar Alto (Spain, 2.2 m), Haleakala T60 (Hawaii, USA), Institut de Radioastronomie Millim\'etrique(IRAM, 30 m), St. Petersburg University LX-200 (40 cm), Kanata telescope (Japan), Nordic Optical Telescope (NOT, La Palma, Spain, 2.56 m), Palomar-Hale telescope (California, USA, 5 m), Boston University Perkins Telescope (1.8 m, Flagstaff, Arizona, USA),  the Sierra Nevada Observatory (1.5 and 0.9 m telescopes, Spain), the Skinakas observatory (Crete, Greece, 1.3 m telescope) and the Submillimeter Array (SMA). The observations and data reduction are described in Appendix \ref{sec:multi_obs}. Figures \ref{plt:opt_pol_lightcurve} and \ref{plt:IR_pol_lightcurve} display the optical and infrared polarized light curves of BL Lac during the {IXPE} observing windows. During the second {IXPE} observation, we were unable to obtain infrared polarization data. During both {IXPE} observations, we find significant variability in polarization degree and angle at millimeter to optical wavelengths.

For {IXPE} obs. 1, the ALMA observations on May 7 yield a median radio polarization degree $\rm \Pi_R=3.95\pm0.3\%$  at 343~GHz along position angle $\psi=23^\circ\pm2^\circ$, and, on May 9, $\rm \Pi_R=3.6\pm0.3\%$ along $\rm \psi_R=33^\circ\pm1^\circ$. The two values of $\rm \Pi_R$ are consistent between scans within the uncertainties. This is also true for $\psi_R$ during the May 7 observation. However, on May 9 we see a change in $\rm \psi_R$ from the first to the final scan from $19^\circ$ to $46^\circ$. The median uncertainty of each measurement is $\pm1^\circ$. The IRAM-30m measurements are the same within the uncertainties, which suggests that there is no significant variability. The median value of $\rm\Pi_R$ at 86~GHz is 4\% with a median uncertainty of 0.4\% along a median position angle $\rm \psi_R=30^\circ\pm3^\circ$. Similarly, at 228.93~GHz, the median $\rm \Pi_R=4.8\pm1.2\%$ along median $\psi=35^\circ\pm7^\circ$. No circular polarization was detected in any of the observations, with a 95\% confidence-interval upper limit of $<$0.44\% and $<$0.86\% for 86~GHz and 228.93~GHz, respectively. We find the optical polarization degree $\rm \Pi_O$ to vary from 1.6\% to 13.7\%,  with a median of 6.8\%. At the same time $\rm \psi_O$ varies from $2^\circ$ to $172^\circ$ with a median of $107^\circ$, almost perpendicular to the jet axis on the plane of the sky ($10^\circ\pm2$, \citealp{Weaver2022}). In the infrared, $\rm \Pi_{\rm IR}$ varies from 0.9\% to 8.4\% with a median of 3.9\%. Note that, although the host galaxy has a negligible contribution to the total emission in the optical during the {\rm IXPE} observation, it is likely that the contribution is much stronger in the infrared. Hence, the $\rm \Pi_{\rm IR}$ measurements should be treated as lower limits to the intrinsic polarization degree. The value of $\rm \psi_{\rm IR}$ varies from $9^\circ$ to $158^\circ$ with a median of $\sim83^\circ$.

During the July {IXPE} observation, the polarization degree detected by the IRAM-30m Telescope decreases at 86~GHz from 8.5\% to 2.2\% with a constant median $\rm \psi_R$ of $\sim15^\circ$, see Fig.~\ref{plt:radio_pol_lightcurve}. The SMA observation at 225~GHz yields $\Pi_R=8.8\pm1\%$ along  $\rm \psi_R$ of $19\pm2^\circ$. The polarization at 228~GHz is consistent within uncertainties at about $\rm \Pi_R=6\%$ along  $\rm \psi_R\approx20^\circ$. At the same time the optical polarization varies from 7\% to 23\% with a median of $\rm \Pi_O=14.2\%$, with $\rm \psi_O$ between $26^\circ$ and $59^\circ$ with a median of $\rm \psi_O=42^\circ$. Tables \ref{tab:mult_obs} and \ref{tab:mult_obs2} summarize the results of the multi-wavelength polarization campaign.

\section{Connections of the X-ray polarization with the Radio and Optical bands}\label{app:restric_tests}

In both leptonic and hadronic models, the X-ray polarimetric properties are tightly related to those at the mm-radio and optical bands, respectively. Motivated by that close connection, we performed additional tests, fixing $\psi_X$ to the corresponding values of $\psi_R$ and $\psi_O$ and computing $\Pi_X$ for the {IXPE} observations.
In a SSC scenario, we expect the polarization angle to be similar to the one of the millimeter-radio seed photons. On the other hand, in hadronic scenarios, the optical polarization degree is expected to be similar to $\Pi_X$. Motivated by these expectations, we proceed to restrict the polarization parameters. We therefore first restrict $\psi_X$ to the value of the mm-radio observations. This seems to improve the $\Pi_X<$ upper limits when fitting the observations separately ($\Pi_X<$ 6.2\% and $\Pi_X<$12.9\% for Obs. 1 and 2, respectively). We repeat the exercise, but this time we restrict $\psi_X$ to the average value of the optical observations. We see a marginal improvement for the second observation with $\Pi_X<$14.2\% and $\Pi_X<$11.9\% for Obs. 1 and 2, respectively. Since $\psi_O$ change by more than 70$^\circ$ from the first to the second IXPE observation we do not attempt a joined fit. The derived upper limits from these tests are also summarized in Table \ref{tab:comparison}. Although improved upper limits as low as $<$6.2\% can be obtained for the first {\it IXPE} observation, none of these attempts significantly enhanced or degraded the fit to the data presented in section \ref{subsec:spec_pol}. In Table~\ref{tab:comparison} we report the corresponding upper limits from our tests.

\begin{table}
\centering
\caption{The upper limits for the X-ray polarization degree $\Pi_X$ derived assuming the X-ray polarization angle to be the same as the average values for $\psi_O$ and $\psi_R$. Upper limits were computed for both the \textit{IXPE} observations.}
\begin{tabular}{lcc}
\hline
Optical angle&$\Pi_X^{\rm Obs1}$ & $\Pi_X^{\rm Obs2}$ \\
\hline
$\psi_O^{\rm Obs1}$=112$^{\circ}$ &$<$14.2\% & \\
$\psi_O^{\rm Obs2}$=38$^{\circ}$ & &$<$11.9\% \\
\hline
Radio angle & &  \\
$\psi_R^{\rm Obs1}$=30$^{\circ}$ &$<$ 6.2\% & \\
$\psi_R^{\rm Obs2}$=18$^{\circ}$ & & $<$12.9\% \\
\hline
\hline
\end{tabular}
\label{tab:comparison}
\end{table}

\section{Discussion \& Conclusions}\label{sec:disc_conc}
We have presented the first X-ray polarization observations of an LBL blazar, BL Lac. The analysis of the {IXPE} data only provides upper limits (corresponding to the 99\% confidence level) to the polarization degree: $\rm \Pi_X<14.2\%$ and $\rm \Pi_X<12.6\%$ for the first and second exposure, respectively. As a consequence, the polarization angle $\rm \psi_X$ is unconstrained for both of the observations. The upper limit to $\rm \Pi_X$ can be decreased to as low as $<$6\% by making assumptions for the $\rm \Pi_X$ and $\rm \psi_X$ only for the first IXPE observation. The upper limits can then be compared with the polarization at longer wavelengths. In the optical, we measure a median $\rm \Pi_O=6.8\%$ at a median angle of $\rm \psi_O=107^\circ$ for the May observation and medians $\rm \Pi_O=14.2\%$ and $\rm \psi_O=42^\circ$ for the July observation. We find evidence of significant mm-radio to optical polarization variability during both {IXPE} observations. Moreover, the variability in the polarization angle is stronger in the optical. Changes in the $\psi$ due to perhaps turbulence or multiple emission regions would reduce the observed polarization degree by $1/\sqrt{N}$, where $N$ is the number of emission regions or turbulent cells \citep{Marscher2014}. Considering the {IXPE} upper limits derived over the observing period, the median $\rm \Pi_O$ during the May observation was a factor of $\sim$2.5 lower, whereas for the July observation $\rm \Pi_O$ was higher than the $\rm \Pi_X$ limit.

A strong synchrotron X-ray component from ultra-high-energy electrons can also occur in BL Lac, but in this case the X-ray spectrum would be much steeper than that of our model fits \citep{Marscher2008}. In a leptonic scenario, under which X-ray and $\gamma$-ray emission arises from Compton scattering, the X-ray polarization degree is expected to be substantially lower than that of the seed photons \citep{Bonometto1973,Nagirner1993,Poutanen1994,Liodakis2019,Peirson2019}. In the case of EC scattering, depending on the scattering geometry and isotropy of the seed photon field, the outgoing radiation could be either unpolarized or polarized.
However, based on previous SED modeling \citep{boettcher2000,boettcher2013,Morris2019,MAGIC2019,Sahakyan2022} we expect SSC emission to dominated over the EC one in the IXPE energy band. In an SSC model, the seed photons are expected to come from mm-radio synchrotron radiation. The mm-radio observations give $\rm \Pi_R\sim4\%$ for the first observation and $\rm \Pi_R\sim6\%$ for the second. This would suggest an expected $\rm \Pi_X$ of $<3\%$ \citep{Peirson2019}. Therefore, our upper limits for any X-ray polarization signal are consistent with a leptonic scenario.
On the other hand, in the case of hadronic processes, X-ray polarization should be less variable, or even stable, compared to  the optical, with negligible depolarization \cite{Zhang2016}. The contribution from synchrotron radiation by protons and secondary particles from collisions involving hadrons is expected to yield a similar, or higher (in the case of a pure proton synchrotron model), value of $\rm \Pi_X$ compared to optical wavelengths \citep{Zhang2013,Paliya2018,Zhang2019}. Alternative emission models involving scattering from relativistic cold electrons are also expected to produce much higher $\rm \Pi_X$ than $\rm \Pi_O$ \citep{Begelman1987}.
During both {IXPE} observations, $\rm \Pi_O$ exceeded the 99\% upper limits of $\rm \Pi_X$ on several occasions. Even considering the median $\rm \Pi_O$ estimates during the {IXPE} observations, the optical still exceeds the X-ray upper limit for the July observation. This difference between the optical and X-ray polarization degrees is in strong tension with the relativistic cold electron scattering model as well as a pure proton-synchrotron model. Although we cannot definitively exclude contribution from hadronic processes to the overall emission, the multiwavelegth polarization observations provide evidence against the hadronic interpretation. Instead, our findings favor leptonic emission, and particularly Compton scattering as the dominant mechanism for the X-ray emission in BL Lac.

{\it \textbf{Acknowledgments:}}
We thank the anonymous referee for her/his bright comments. The Imaging X-ray Polarimetry Explorer (IXPE) is a joint US and Italian mission.  The US contribution is supported by the National Aeronautics and Space Administration (NASA) and led and managed by its Marshall Space Flight Center (MSFC), with industry partner Ball Aerospace (contract NNM15AA18C).  The Italian contribution is supported by the Italian Space Agency (Agenzia Spaziale Italiana, ASI) through contract ASI-OHBI-2017-12-I.0, agreements ASI-INAF-2017-12-H0 and ASI-INFN-2017.13-H0, and its Space Science Data Center (SSDC), and by the Istituto Nazionale di Astrofisica (INAF) and the Istituto Nazionale di Fisica Nucleare (INFN) in Italy.  This research used data products provided by the IXPE Team (MSFC, SSDC, INAF, and INFN) and distributed with additional software tools by the High-Energy Astrophysics Science Archive Research Center (HEASARC), at NASA Goddard Space Flight Center (GSFC). We acknowledge financial support from ASI-INAF agreement n.\ 2022-14-HH.0. The research at Boston University was supported in part by National Science Foundation grant AST-2108622 and NASA Swift Guest Investigator grant 80NSSC22K0537. This research has made use of data from the RoboPol programme, a collaboration between Caltech, the University of Crete, IA-FORTH, IUCAA, the MPIfR, and the Nicolaus Copernicus University, which was conducted at Skinakas Observatory in Crete, Greece. The IAA-CSIC co-authors acknowledge financial support from the Spanish "Ministerio de Ciencia e Innovacion" (MCINN) through the "Center of Excellence Severo Ochoa" award for the Instituto de Astrof\'{i}sica de Andaluc\'{i}a-CSIC (SEV-2017-0709). Acquisition and reduction of the POLAMI, TOP-MAPCAR, and OSN data was supported in part by MICINN through grants AYA2016-80889-P and PID2019-107847RB-C44. The POLAMI observations were carried out at the IRAM 30m Telescope. IRAM is supported by INSU/CNRS (France), MPG (Germany) and IGN (Spain). This paper makes use of the following ALMA director's discretionary time data under proposal ESO\#2021.A.00016.T. ALMA is a partnership of ESO (representing its member states), NSF (USA) and NINS (Japan), together with NRC (Canada), MOST and ASIAA (Taiwan), and KASI (Republic of Korea), in cooperation with the Republic of Chile. The Joint ALMA Observatory is operated by ESO, AUI/NRAO and NAOJ. Some of the data reported here are based on observations obtained at the Hale Telescope, Palomar Observatory as part of a continuing collaboration between the California Institute of Technology, NASA/JPL, Yale University, and the National Astronomical Observatories of China. This research made use of Photutils, an Astropy package for detection and photometry of astronomical sources (Bradley et al., 2019). GVP acknowledges support by NASA through the NASA Hubble Fellowship grant  \#HST-HF2-51444.001-A awarded by the Space Telescope Science Institute, which is operated by the Association of Universities for Research in Astronomy, Inc., under NASA contract NAS5-26555. The data in this study include observations made with the Nordic Optical Telescope, owned in collaboration by the University of Turku and Aarhus University, and operated jointly by Aarhus University, the University of Turku and the University of Oslo, representing Denmark, Finland and Norway, the University of Iceland and Stockholm University at the Observatorio del Roque de los Muchachos, La Palma, Spain, of the Instituto de Astrofisica de Canarias. The data presented here were obtained in part with ALFOSC, which is provided by the Instituto de Astrof\'{\i}sica de Andaluc\'{\i}a (IAA) under a joint agreement with the University of Copenhagen and NOT. E.\ L.\ was supported by Academy of Finland projects 317636 and 320045. Part of the French contribution is supported by the Scientific Research National Center (CNRS) and the French Spatial Agency (CNES). Some of the data are based on observations collected at the Observatorio de Sierra Nevada, owned and operated by the Instituto de Astrof\'{i}sica de Andaluc\'{i}a (IAA-CSIC). Further data are based on observations collected at the Centro Astron\'{o}mico Hispano-Alem\'{a}n(CAHA), operated jointly by Junta de Andaluc\'{i}a and Consejo Superior de Investigaciones Cient\'{i}ficas (IAA-CSIC). D.B., S.K., R.S., N. M., acknowledge support from the European Research Council (ERC) under the European Unions Horizon 2020 research and innovation programme under grant agreement No.~771282. CC acknowledges support by the European Research Council (ERC) under the HORIZON ERC Grants 2021 programme under grant agreement No. 101040021. The Dipol-2 polarimeter was built in cooperation by the University of Turku, Finland, and the Leibniz Institut f\"{u}r Sonnenphysik, Germany, with support from the Leibniz Association grant SAW-2011-KIS-7. We are grateful to the Institute for Astronomy, University of Hawaii, for the allocated observing time. A. H. acknowledges The National Radio Astronomy Observatory is a facility of the National Science Foundation operated under cooperative agreement by Associated Universities, Inc.’ This work was supported by JST, the establishment of university fellowships towards the creation of science technology innovation, Grant Number JPMJFS2129. This work was supported by Japan Society for the Promotion of Science (JSPS) KAKENHI Grant Numbers JP21H01137. This work was also partially supported by Optical and Near-Infrared Astronomy Inter-University Cooperation Program from the Ministry of Education, Culture, Sports, Science and Technology (MEXT) of Japan. We are grateful to the observation and operating members of Kanata Telescope. The Submillimeter Array is a joint project between the Smithsonian Astrophysical Observatory and the Academia Sinica Institute of Astronomy and Astrophysics and is funded by the Smithsonian Institution and the Academia Sinica. Maunakea, the location of the SMA, is a culturally important site for the indigenous Hawaiian people; we are privileged to study the cosmos from its summit.

\facilities{ALMA, AZT-8, Calar Alto, IRAM-30m, IXPE, LX-200, Kanata, Nordic Optical Telescope, NuSTAR, Palomar, Perkins, Skinakas Observatory, SMA, Swift, T60, T90, T150, XMM-Newton}

\appendix

\section{Multi-wavelength observations}\label{sec:multi_obs}

\subsection{mm-radio observations}
\begin{figure*}
\centering
\includegraphics[width=0.99\textwidth]{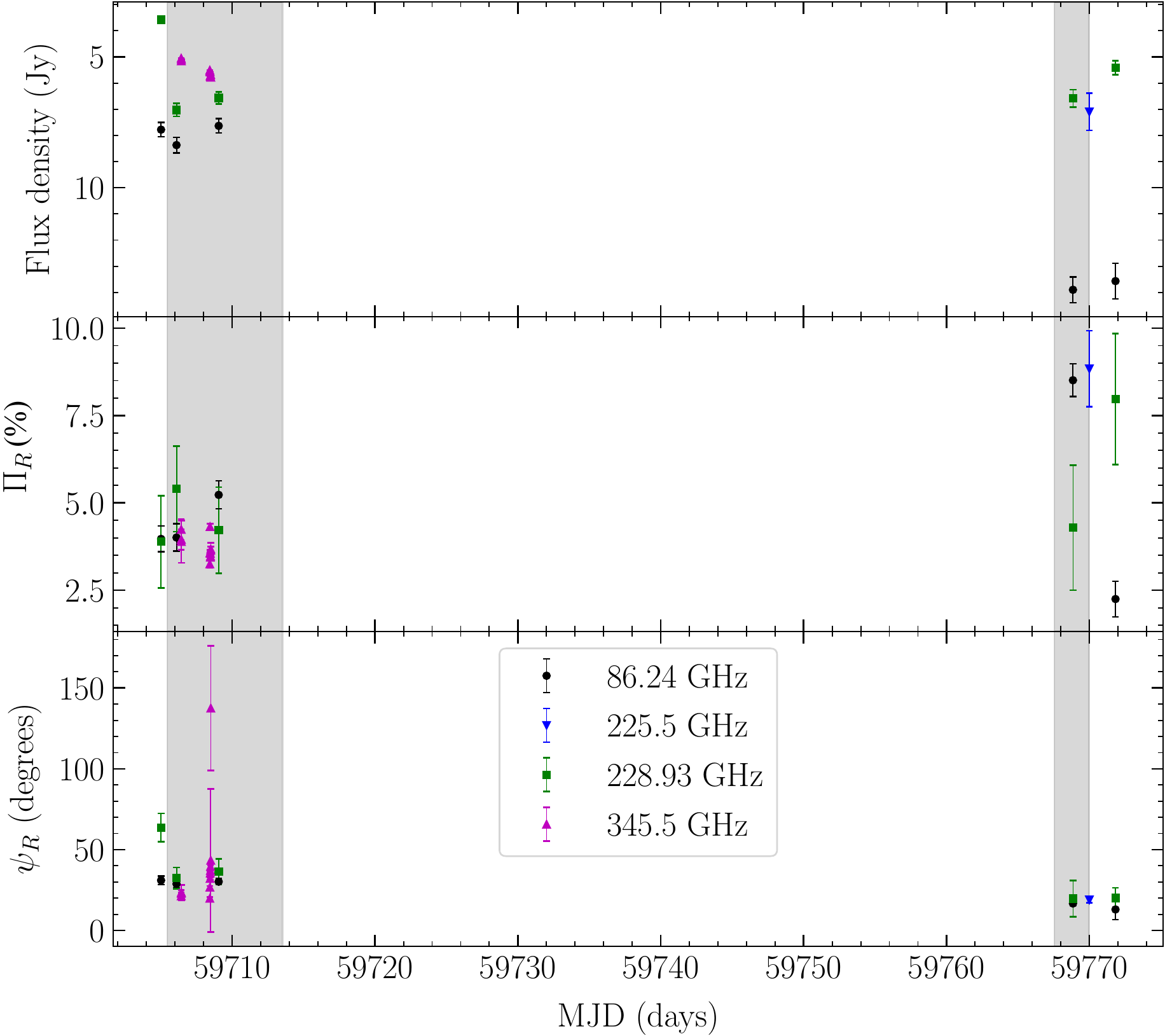}
\caption{Radio polarization vs.\ time for BL Lac. {\it Top:} flux density {\it middle:} polarization degree, {\it bottom:}  polarization angle. The gray shaded areas demark the duration of {IXPE} observations 1 and 2.}
\label{plt:radio_pol_lightcurve}
\end{figure*}

Radio polarization observations were obtained at millimeter and sub-millimeter wavelengths using the Atacama Large Millimeter/sub-millimeter Array (ALMA), the Institut de Radioastronomie Millim\'{e}trique 30-m Telescope (IRAM-30m), and the Submillimeter Array (SMA). The ALMA observations were obtained in band 7 (mean wavelength of 0.87 mm, frequency 345 GHz) on 2022 May 7 and 9. The ALMA observations were reduced using the AMAPOLA\footnote{\url{http://www.alma.cl/~skameno/AMAPOLA/}} polarization pipeline, which is used to estimate polarization properties from short monitoring observations of ALMA grid sources for calibrator selection. It aims to determine the D-terms for instrumental calibration from short scans of different sources with sufficient S/N and employs an antenna-based database for a-priori values assuming stability. Thus the method is applicable for observations when less than 60 degrees of field rotation (parallactic angle) is achieved on the polarization calibrator. The standard reduction procedure assumes a larger parallactic angle coverage for the determination of the D-terms on the polarization calibrator, which are then transferred instantaneously for the calibration of the target\footnote{\url{http://www.alma.cl/~skameno/POLBEAM/ShortCalibrationSchemeTests20170321.pdf},\url{http://www.alma.cl/~skameno/POL/ShortPOL/ShortPolCal20170830.pdf}}$^,$\footnote{\url{http://www.alma.cl/~skameno/POL/ShortPOL/ShortPolCal20170830.pdf}}. The SMA observation was obtained within the SMAPOL monitoring program on 10 July 2022 at 1.3 mm corresponding to 225.538~GHz. The IRAM-30m observations were performed on 2022 May 5, 7, and 10, and again on 2022 July 8 and 11 at 3.5 mm (86.24 GHz) and 1.3 mm (228.93 GHz) as part of the IRAM's Polarimetric Monitoring of AGN at Millimeter Wavelengths (POLAMI) Large Project\footnote{\url{http://polami.iaa.es/}} \citep{Agudo2018, Agudo2018-II, Thum2018}. In Fig.~\ref{plt:radio_pol_lightcurve} we show the mm-radio polarization light curve.

\subsection{Optical and infrared observations}

\begin{figure*}
\centering
\includegraphics[width=0.99\textwidth]{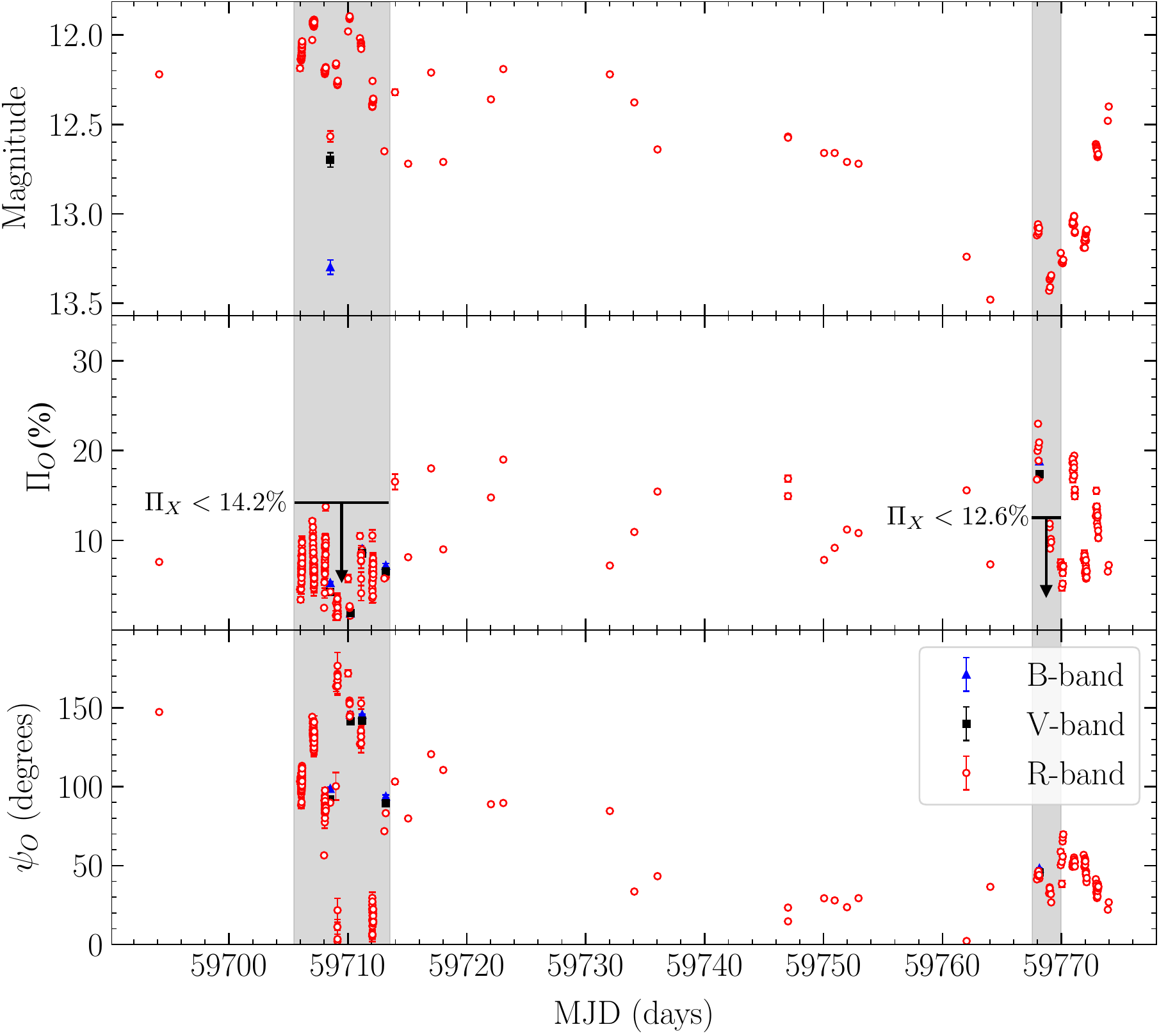}
\caption{Optical polarization time variations for BL Lac. {\it Top:} magnitude, {\it middle:} polarization degree, {\it bottom:} polarization angle. The gray shaded areas demark the duration of the {IXPE} observations. The IXPE upper limits obtained in the two observations are also reported.}
\label{plt:opt_pol_lightcurve}
\end{figure*}

\begin{figure*}
\centering
\includegraphics[width=0.99\textwidth]{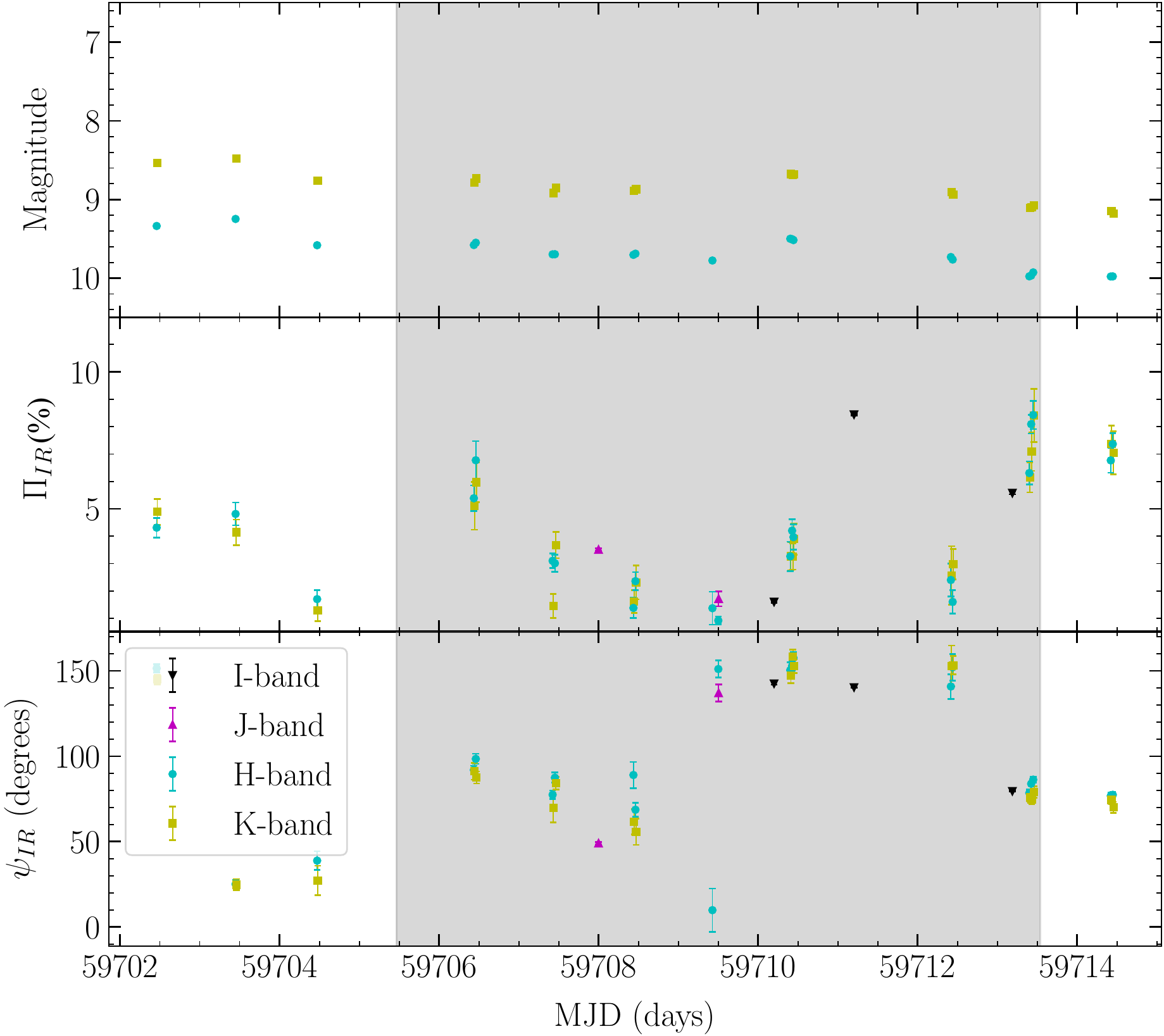}
\caption{Infrared polarization vs.\ time for BL Lac. {\it Top:} magnitude {\it middle:} polarization degree, {\it bottom:}  polarization angle. The gray shaded area demarks the duration of {IXPE} observation 1.}
\label{plt:IR_pol_lightcurve}
\end{figure*}

During {IXPE} observation 1, BL Lac was observed in optical polarization by the AZT-8 telescope (2022 May 6-11), the Calar Alto observatory (2022 May 7, 9, 11, and July 7-9), the Haleakala observatory T60 telescope (2022 May 9), Kanata telescope (2022 May 9), the Nordic Optical Telescope (NOT, 2022 May 11, 12, and 14 and 2022 July 7), Palomar observatory (Hale, 2022 May 10), the Perkins observatory (2022 May 3-5, 7-11, 13-15), the St. Petersburg University LX-200 telescope  (2022 May 6, 9, 14), the Sierra Nevada observatory (T90 and T150, 2022 May 7-13, and July 9 ) and the Skinakas observatory (RoboPol, 2022 May 14, 16, and July 7, 9). 
The Calar Alto Observatory observations used the 2.2 m telescope and the imaging polarimetric mode of the Calar Alto Faint Object Spectrograph (CAFOS). Observations were obtained in the $R_c$ filter and reduced using both unpolarized and polarized standards stars and following standard analysis procedures. Similar procedures and the same filter was used for the T90 and T150 telescope observations at the Sierra Nevada Observatory. We performed R-band polarimetric observations with the Hiroshima Optical and Near-InfraRed camera (HONIR, \citealp{Akitaya2014}) installed on the Kanata telescope. The polarization degree, polarization angle, and corresponding errors were estimated from Stokes parameters obtained from four exposures at positions 0$^\circ$, 45$^\circ$, 22.5$^\circ$, and 67.5$^\circ$ of the half-wave plate for each observation \citep{Kawabata1999}. Offset angle and wiregrid depolarization were corrected using highly polarized standard stars (BD+64d106, BD+59d389). The instrumental polarization was determined with the help of unpolarized standard stars (HD14069) to be $<0.2\%$. We also obtained linear polarimetric observations with the Alhambra Faint Object Spectrograph and Camera (ALFOSC) in $B,~V,~R,~I$ bands of BL Lac, along with polarized and unpolarized standard stars during each of the observing nights for instrumental calibration. The data were reduced following standard photometric procedures included in the Tuorla Observatory's data reduction pipeline, described in detail in \cite{Hovatta2016} and \cite{Nilsson2018}. The T60 telescope uses the "double-image" CCD polarimeter Dipol-2 \citep{Piirola2014}. Dipol-2 is capable of simultaneously observing in $B$, $V$, and $R$ filters \citep{Piirola1973,Berdyugin2018,Berdyugin2019,Piirola2021}. The instrumental polarization and zero point of the polarization angle were determined by observing polarized and unpolarized standard stars, and the measurements are combined using the ``2 $\times$ sigma-weighting algorithm''. The standard error of the weighted means of the normalized Stokes parameters are then propagated to obtain the final uncertainty of the polarization degree and angle \citep{Kosenkov2017,Piirola2021}. The Skinakas observatory observations used the RoboPol instrument mounted in the 1.3-m telescope \citep{Ramaprakash2019}. RoboPol is a novel 4-channel polarimeter that simultaneous measures the normalized Stokes $q$ and $u$ parameters with a single exposure and no moving parts. The data reduction and analysis pipeline is described in detail in \cite{Panopoulou2015} and \cite{Blinov2021}.  The 40cm LX-200 and 70cm AZT-8 telescopes are equipped with nearly identical imaging photo-polarimeters based on a ST-7 camera, using and swapping two Savart plates oriented 45$^\circ$ with respect to each other. The observations were performed in the $R$-band and the data were background, bias, and flat field corrected, as well as instrumental and interstellar polarization calibrated with the use of standard stars.

In addition to the optical measurements, we obtained observations in the $J$, $H$, and $K$ infrared bands using the 200-inch Palomar Hale telescope, the Kanata telescope, and the WIRC+Pol instrument \citep{Tinyanont2019a}. The Hale telescope observations were performed in the $J$ and $H$ bands using a polarized grating to simultaneously measure four linearly polarized components, while a half-wave plate improved polarimetric sensitivity by beam-swapping \citep{Tinyanont2019b,Millar-Blanchaer2021}. The data were reduced using the WIRC+Pol Data Reduction Pipeline software\footnote{\url{https://github.com/WIRC-Pol/wirc_drp}}, described in detail in  \cite{Tinyanont2019a} and \cite{Masiero2022}. The Kanata observation was performed in the J-band simultaneously to the R-band observation (see above). Data from the Perkins Telescope were obtained in the $H$, and $K$ bands, using the IR camera MIMIR\footnote{https://people.bu.edu/clemens/mimir/index.html}. One measurement consists of 6 dithering exposures of 3 s each at 16 positions of a half-wave plate, rotated in steps of 22.5\degr~from 0 to 360\degr. The camera and data reduction are described in detail in \cite{Clemens2012}.

\subsection{X-ray observations}\label{app:X-ray_obs}

		\begin{table}
		\centering
		\caption{\small{Log of X-ray observations related to the IXPE pointings of BL Lac. The log of the \textit{Swift} exposures and the corresponding derived quantities are reported in Appendix \ref{sec:swift_obs}.}}\label{log}
		\begin{tabular}{c c c c}
			\hline
			Observatory & Obs. ID & Obs. date & Net exp. \\
			& & yyyy-mm-dd & ks \\
			\hline 
			{IXPE} & 01006301  & 2022/05/06-14 &$\sim$ 390\\
			{NuSTAR} & 60701036002  & 2022/5/6 & $\sim$ 22\\
			{XMM-Newton} & 0902111001& 2022/5/15&$\sim$7\\
			{IXPE} & 01006701  & 2022/7/7-9 &$\sim$ 116\\
			{XMM-Newton} & 0902111301& 2022/7/8-9&$\sim$15\\
            \hline 
		\end{tabular}
	\end{table}
	
We here present the data reduction of the different datasets studies in Sect. 2 and obtained using {IXPE, XMM-Newton}. In Table~\ref{log}, we summarize their corresponding information.

For both IXPE observations, the cleaned event files and the associated science products were obtained using a dedicated pipeline relying on the \textit{Ftools} software package and adopting the latest calibration data files from {IXPE} (CALDB 20211118). The source regions for each of the three detector units (DUs) were then selected via an iterative process aimed at maximizing the signal-to-noise ratio (SNR) in the {IXPE} operating energy range of 2--8 keV. In particular, we used circular regions with radius $47''$ for all three DUs. A constant energy binning of 7 counts per bin was used for $Q$ and $U$ stokes parameters; we required SNR$>7$ in each channel for the intensity spectra. We then performed a so-called weighted analysis method presented in \citet{dimarco2022} (parameter {\sc stokes=Neff} in {\sc xselect}) on the resulting spectra. We adopted a circular region with radius 104\arcsec~to determine the $I$, $Q$, and $U$ Stokes background spectra.

The {XMM-Newton} scientific products were obtained with the standard {SAS} routines and the latest calibration files. The spectrum of the source was derived using a circular region (radius$=40''$) centered on the source. The background was extracted from a blank region on the {Epic-pn} CCD camera using a circular region of the same size. The resulting spectrum was rebinned in order to have at least 30 counts in each bin and to avoid oversampling the spectral resolution by a factor $>3$. The {NuSTAR} data were calibrated and cleaned using the {NuSTAR} Data Analysis Software \citep[NuSTARDAS\footnote{\url{https://heasarc.gsfc.nasa.gov/docs/nustar/analysis/nustar_swguide.pdf}},][]{Perri2021}, and the scientific products were generated with the {\it nuproducts} pipeline using the latest calibration database (v. 20220302). The source spectrum was derived using a circular region (radius $=70''$), and a concentric annulus (r$_{\rm in}$ and r$_{\rm out}$ being $270''$ and $370''$, respectively) was used to derive the background spectrum.

\begin{table}
\centering
\caption{Multi-wavelength polarization observations related to the May 2022 {\rm IXPE} pointing. }
\begin{tabular}{lccccc}
\hline
Telescope & X-ray flux & $\rm \Pi_X$ (\%)& $\sigma_{\Pi}$ & $\rm \psi_X$ (deg) & $\sigma_{\psi}$\\
\hline
{IXPE}& $0.96\pm0.03(0.05)$ & $<$14.2 & - & - & -\\
{XMM-Newton}& $0.91\pm0.02(0.03)$ & - & - & - & -\\
{NuSTAR} &$1.14\pm0.03(0.05)$ & - & - & - & -\\
\hline
\hline
Telescope & Flux density (Jy) & $\rm \Pi_R$  (\%)& $\sigma_{\Pi}$ & $\rm \psi_R$ (deg.) & $\sigma_{\psi}$ \\
\hline
ALMA (0.87 mm) & $5.58\pm0.26$ &  $3.6\pm0.09$ & 0.3 & $26\pm0.9$ & 31\\
POLAMI (3 mm) & $7.78\pm0.31$ &  $4.0\pm0.39$  & 0.6 & $30\pm3$ & 1\\
POLAMI (1.3 mm) & $6.57\pm1.5$   &  $4.2\pm1.2$ & 0.6 & $36\pm7.9$ & 13\\
\hline
\hline
Telescope & Magnitude & $\rm \Pi_O$ (\%)& $\sigma_{\Pi}$ &  $\rm \psi_O$ (deg) & $\sigma_{\psi}$\\
\hline 
AZT-8 \& LX-200 (70 and 40 cm)& $12.8\pm0.1$ & $5.0\pm0.3$ &3.45 & $116\pm2$ & 28\\
Calar Alto & $12.8\pm0.1$ & $6.8\pm0.1$ &2.8 & $128\pm1$ &23 \\
Kanata (R-band)& -- & $2.49\pm0.05$ & 0 &$56.5\pm0.5$ & 0\\
Kanata (J-band)& -- & $3.51\pm0.05$ & 0 &$49.0\pm0.6$  & 0\\
NOT & --  & $5.9\pm3.0$ &3.0 & $140\pm28$ &28 \\
Palomar (J-band) & -- & $1.70\pm0.27$ &0 & $137\pm5$&0\\
Palomar (H-band) & -- & $0.92\pm0.15$ &0 & $151\pm5$&0\\
Perkins (H-band) & $9.69\pm0.2$ & $4.0\pm0.4$ &2.3 & $86\pm3$ &38\\
Perkins (K-band) & $8.87\pm0.2$ & $3.4\pm2.1$ &2.1 & $79\pm4$ &37\\
Sierra Nevada (T150) & $12.8\pm0.2$ & $6.9\pm0.6$ &1.8 & $100\pm3$&46 \\
Skinakas& $12.68\pm0.03$ & $5.77\pm0.1$ &0 & $71\pm0.5$ &0\\
T60 & $13.28\pm0.03$ & $4.28\pm0.09$ &0 & $90\pm1$ & 0\\
\hline
\end{tabular}
\tablecomments X-ray fluxes correspond to the 2--8 keV energy range and are in units of 10$^{-11}$ erg cm$^{-2}$ s$^{-1}$. The mm-radio flux density is in Janskys (Jy). For the optical observations, we report R-band measurements. The infrared observations are affected by the unpolarized host-galaxy contribution to the total light, and so should be treated as lower limits to the true $\Pi$. The uncertainties for $\Pi$ and $\psi$ are either the  uncertainty of the measurement or the median uncertainty in the case of multiple measurements. $\sigma_{\Pi}$ and $\sigma_{\psi}$ show the standard deviation of the observations.
\label{tab:mult_obs}
\end{table}

\begin{table}
\centering
\caption{Multi-wavelength polarization observations related to the July 2022 {IXPE} observation.}
\begin{tabular}{lccccc}
\hline
Telescope & X-ray flux & $\rm \Pi_X$ (\%)&$\sigma_\Pi$ &  $\rm\psi_X$ (deg.) & $\sigma_\psi$\\
\hline
{IXPE}& $1.56\pm0.06(0.09)$ & $<12.6$ & - & - & -\\
{XMM-Newton}& $1.60\pm0.01(0.02)$ & - & - & - & -\\
\hline
\hline
Telescope & Flux density (Jy) & $\rm\Pi_R$ (\%)& $\sigma_\Pi$ & $\rm \psi_R$ (deg.)& $\sigma_\psi$\\
\hline
POLAMI (3 mm) & $13.7\pm0.6$ & $5.4\pm0.5$& 3.1 & $15\pm4$ & 2\\
POLAMI (1.3 mm) & $6.0\pm0.6$ & $6.1\pm1.8$ & 1.8 & $20\pm9$& 0.2\\
SMA (1.3 mm) & $7.0\pm0.7$ & $8.8\pm1.0$ & 0 & $19\pm2$ & 0\\
\hline
\hline
Telescope & Magnitude & $\rm \Pi_O$ (\%)& $\sigma_\Pi$&  $\rm \psi_O$ (deg.) & $\sigma_\psi$\\
\hline 
Calar Alto & $13.9\pm0.1$  & $15.9\pm0.1$& 5.2 & $40\pm0.1$ &7 \\
NOT& -- & $17.0\pm0.08$ & 0 & $42\pm4$ & 0\\
Sierra Nevada (T90) & $13.9\pm0.01$ & $7.3\pm0.36$ & 0.2 & $54\pm1.4$&4 \\
Skinakas& $13.26\pm0.15$ & $13.2\pm0.1$ & 4.2& $38\pm0.3$& 5\\
\hline
\hline
\end{tabular}
\tablecomments{Same as for Table \ref{tab:mult_obs}}
\label{tab:mult_obs2}
\end{table}

\section{Swift-XRT observations: temporal behavior of BL Lac}\label{sec:swift_obs}
\begin{figure*}
    \centering
    \includegraphics[width=\textwidth]{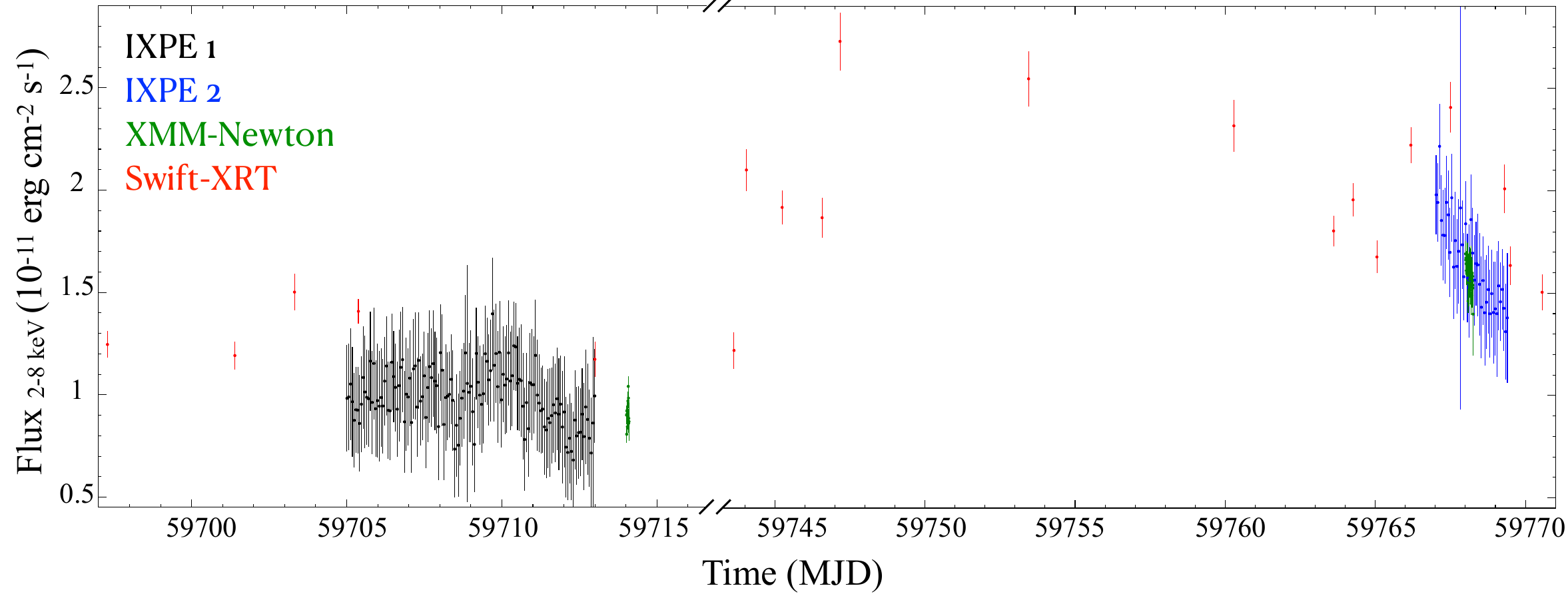}
    \caption{Multi-mission light-curve of BL Lac as observed in the 2--8 keV energy range. Different colors account for the various facilities. No significant intra-observation variability is observed during the first {IXPE} pointing, while, a flux decrease by a factor of $\sim$30\% is observed during the second exposure.}
    \label{plt:multilc}
\end{figure*}

We here report on a list the {Swift-XRT} exposures that were obtained in the context of a monitoring campaign aimed at tracking the flux level of BL Lac. Scientific products from the {Swift-XRT} exposures were 
derived using the facilities provided by the Space Science Data Center (SSDC\footnote{\url{https://www.ssdc.asi.it/}}) of the Italian Space Agency (ASI). In particular, the source spectra were extracted with a circular region of radius $\sim47''$, with a concentric annulus for determination of the background with inner(outer) radii of 120(150) arcseconds. The spectra were then binned in order to include at least 25 counts in each bin. We modeled each of the obtained 21 {XRT} spectra as a simple power-law with Galactic photoelectric absorption. This model was found to adequately reproduce the data, based on the $\chi^2$ statistic. We report the 2--8 keV fluxes and the inferred photon indices in Table \ref{xrtlog}. We then used \textit{Swift} light curve to study the variability properties of BL Lac over $\sim$3 months preceding and including the dates in which {IXPE} was observing BL Lac. The photon index as well as the 2--8 keV flux of BL Lac were derived for each of the {XRT} exposure fitting a simple power law observed for the Galaxy. Our result, quoted in Table \ref{tab:swift_obs}, are in agreement with a harder when brighter behavior as the 2--8 flux and $\Gamma$ are moderately anti-correlated with a Pearson cross-correlation coefficient of $P_{ \rm cc}=-0.6$ and an accompanying null probability $P(<r)=0.004$. This behavior has been already observed in blazars and BL Lac itself \citep[e.g.][]{Prince2021} and is suggestive of a non-flaring activity of the source. Interestingly, this anti-correlation is moderately stronger if we consider the 2--10 keV flux ($P_{\rm cc}=$--0.64 and $P(\rm <r)=0.002$), while no relation between the photon index and 0.5--2 keV flux is found in this dataset.
Finally, in Fig. ~\ref{plt:multilc}, we report the flux variability and compare it with the two {IXPE} and {XMM-Newton} light-curves. 

\begin{table}
\centering
\caption{The {Swift-XRT} observations belonging to the BL Lac monitoring campaign before, during and after the two {IXPE} observations.}\label{tab:swift_obs}
\begin{tabular}{lcccc}
\hline
Start Time & Time & ObsID  &  Flux$_{\rm 2-8~keV}$ & $\Gamma$\\
yyyy-mm-dd hh:mm:ss  & (MJD) &  & ($\rm 10^{-11} erg cm^{-2}$ s$^{-1}$) & \\
\hline
2022-04-28 06:55:35&59697.288&00096565001&1.24$\pm$0.06&1.54$\pm$0.15\\
2022-05-02 09:27:36&59701.394&00096565002&1.19$\pm$0.06&1.69$\pm$0.15\\
2022-05-04 07:36:35&59703.317&00096565003&1.50$\pm$0.09&1.56$\pm$0.16\\
2022-05-06 09:01:35&59705.376&00089271001&1.40$\pm$0.15&1.57$\pm$0.11\\
2022-05-13 23:57:35&59712.998&00096565004&1.17$\pm$0.08&1.36$\pm$0.22\\
2022-06-12 18:40:35&59742.778&00014925008&1.35$\pm$0.09&1.56$\pm$0.18\\
2022-06-13 15:21:45&59743.640&00014925009&1.22$\pm$0.08&1.53$\pm$0.22\\
2022-06-14 01:16:35&59744.053&00014925010&2.10$\pm$0.10&1.25$\pm$0.17\\
2022-06-15 05:50:35&59745.243&00014925011&1.91$\pm$0.08&1.45$\pm$0.13\\
2022-06-16 13:47:34&59746.574&00014925012&1.86$\pm$0.09&1.33$\pm$0.17\\
2022-06-17 04:04:35&59747.169&00014925013&2.73$\pm$0.14&1.44$\pm$0.15\\
2022-06-23 10:51:36&59753.452&00096990001&2.54$\pm$0.13&1.26$\pm$0.16\\
2022-06-30 06:56:36&59760.289&00096990002&2.31$\pm$0.12&1.49$\pm$0.16\\
2022-07-03 14:38:35&59763.610&00096990003&1.80$\pm$0.07&1.35$\pm$0.13\\
2022-07-04 06:13:38&59764.259&00096990004&1.96$\pm$0.08&1.45$\pm$0.12\\
2022-07-05 01:17:37&59765.054&00096990005&1.67$\pm$0.07&1.36$\pm$0.15\\
2022-07-06 04:23:36&59766.183&00096990006&2.22$\pm$0.08&1.40$\pm$0.12\\
2022-07-07 12:09:37&59767.506&00096990007&2.40$\pm$0.12&1.28$\pm$0.17\\
2022-07-09 07:22:36&59769.307&00096990008&2.01$\pm$0.11&1.36$\pm$0.19\\
2022-07-09 11:54:36&59769.496&00096990009&1.63$\pm$0.09&1.55$\pm$0.16\\
2022-07-10 13:21:36&59770.556&00096990010&1.50$\pm$0.08&1.55$\pm$0.16\\
\hline
\end{tabular}
\label{xrtlog}
\end{table}

\bibliography{biblio}
\bibliographystyle{aasjournal}

\end{document}